\documentclass[10pt,twocolumn,superscriptaddress,showpacs,amsmath,amssymb,aps,pra,floatfix]{revtex4-1}

\usepackage{float}
\usepackage{lipsum}
\usepackage{epsfig}
\usepackage{amsmath}
\usepackage{amssymb}
\usepackage{bm}
\usepackage{graphicx}
\usepackage{color}
\usepackage{graphicx,rotating}

\begin{document}

\title{Electromagnetic fields of an ultra-short tightly-focused radially-polarized laser pulse}

\author{Yousef I. Salamin}
\email{ysalamin@aus.edu}
\affiliation{Max-Planck-Institut f\"{u}r Kernphysik, Saupfercheckweg 1,
D-69029 Heidelberg, Germany}
\affiliation{Department of Physics, American University of Sharjah, POB 26666, Sharjah, United Arab Emirates}

\author{Jian-Xing Li}
\email{Jian-Xing.Li@mpi-hd.mpg.de}
\affiliation{Max-Planck-Institut f\"{u}r Kernphysik, Saupfercheckweg 1,
D-69029 Heidelberg, Germany}

\date{\today}

\begin{abstract}

Fully analytic expressions, for the electric and magnetic fields of an ultrashort and tightly focused laser pulse of the radially polarized category, are presented to lowest order of approximation. The fields are derived from scalar and vector potentials, along the lines of our earlier  work for a similar pulse of the linearly polarized variety. A systematic program is also described from which the fields may be obtained to any desired accuracy, analytically or numerically.

\end{abstract}

\maketitle

\section{Introduction}

Present-day high-power laser systems, and ones that are currently envisaged for the near future \cite{eli, hiper,xfel}, are mostly of the pulsed type. For many applications \cite{marklund, mourou, esarey2009, piazza, jianxing2014, krushelnick, kitagawa, jianxing2015}, including laser acceleration of electrons, ions and bare nuclei, and laser-matter interaction, the need for super-intense pulses which contain merely a few laser cycles, continues to grow. Hence, a proper theoretical representation for the electric and magnetic fields of a pulse of this type is very much in demand and continues to motivate work, and stimulate efforts, in the field \cite{lax,davis,barton, esarey, esarey2, porras, brabec, winful,wang2,lu,hua,yan,becker,sepke_pulse,varin,pukhov, april1,april2, gonoskov,marceau, wong, varin1, salamin92a,salamin92b,josab33}.

A radially-polarized laser beam or pulse has two electric field components, one radial, which helps to confine a beam of particles to a region close to the propagation axis, and a second that is axial, which comes in handy for particle acceleration \cite{salamin2010,jianxing2012, wong2, sell}. This work is also motivated by recent advances in the technology of radially polarized laser systems \cite{kaertner, carbajo}. For modeling ultra-short ultra-strong laser pulses, the paraxial solutions of Maxwell's equations are no longer adequate \cite{lax, davis, barton}. The need for more accurate solutions to be employed \cite{esarey, april2} seems timely now.

This paper contributes to those continuing efforts and builds upon former investigations \cite{salamin92a,josab33,salamin92b}. Analytic expressions for the most dominant terms in the description of the electric and magnetic fields of an ultra-short and tightly-focused laser pulse of the radially-polarized category, will be derived. For our purposes in this work, ultra-short will mean of axial length, $L$, small compared to a Rayleigh length $z_r$, and by tightly-focused will be meant of waist radius at focus, $w_0\leq\lambda_0$, where $\lambda_0$ is a central wavelength.

This paper is organized as follows. First, key points of the background material are briefly reviewed in Sec. \ref{sec:back}. Based on that, analytic expressions for the zeroth-order fields, in some truncated series, assumed to be the most dominant ones, will be derived in Sec. \ref{sec:EandB}, following the work of Esarey {\it et al.} \cite{esarey} and our own earlier investigations \cite{salamin92a,josab33}. Higher-order corrections will only be used in numerical simulations, as they turn out to be quite cumbersome. In the derivations, two different initial pulse frequency-distributions will be employed, one Gaussian and the other Poissonian. Two sets of zero-order fields, obtained from the two distributions, will be derived and compared analytically, as well as numerically. A summary and our main conclusions will be given in Sec. \ref{sec:conc}.

\section{Background}\label{sec:back}

The fields will be derived from a vector potential $\bm{A}$ which satisfies the wave equation
\begin{equation}\label{we}
  \bm{\nabla}^2\bm{A} - \frac{1}{c^2} \frac{\partial^2\bm{A}}{\partial t^2} = 0,
\end{equation}
and a similar equation for the scalar potential $\Phi$ in vacuum, where $c$ is the speed of light. The vector and scalar potentials will be assumed to be linked via a Lorentz condition. The Following is a brief outline of the steps leading from Eq. (\ref{we}) to an expression for $\bm{A}$, and the associated scalar potential, from which the $\bm{E}$ and $\bm{B}$ fields of a radially polarized laser pulse will ultimately be obtained \cite{salamin92a}. The starting point is a change of variables, to the set $(\rho,\eta,\zeta)$, where $\rho = r/w_0$, $r=\sqrt{x^2+y^2}$, $w_0$ is the initial waist radius at focus, $\zeta=z-ct$, and $\eta=(z+ct)/2$. In terms of the new variables, and for propagation along the $z$-axis, the ansaz
\begin{equation}\label{ansaz}
  A = A_0 a(\rho,\eta,\zeta) e^{ik_0\zeta},
\end{equation}
for the single-component vector potential is then introduced, in which $A_0$ is a constant complex amplitude, and $k_0=2\pi/\lambda_0$ is a central wavenumber corresponding to the central wavelength $\lambda_0$. The coordinate transformations turn (\ref{we}) into an equation for the amplitude $a(\rho,\eta,\zeta)$, which may then be synthesized from Fourier components $a_k$, according to
\begin{equation}\label{ak}
  a(\rho,\eta,\zeta) = \frac{1}{\sqrt{2\pi}}\int_{-\infty}^\infty a_k(\rho,\eta,k) e^{ik\zeta} dk.
\end{equation}
It can be easily shown that each Fourier component satisfies an equation of the form
\begin{equation}\label{akwe}
\left[\frac{1}{\rho}\frac{\partial}{\partial\rho}\rho\frac{\partial}{\partial\rho} +4iz_{rk}\frac{\partial}{\partial \eta}\right]a_k = 0;
	\quad z_{rk} = (k+k_0)\frac{w_0^2}{2}.
\end{equation}
Equation (\ref{akwe}) admits an exact analytical solution, which may be written as
\begin{equation}\label{aksol}
  a_k(\rho,\eta,k) = f_k\psi_k,
\end{equation}
where
\begin{equation}\label{psik}
  \psi_k = \beta_ke^{-\beta_k\rho^2};\quad \beta_k=\frac{1}{1+i\alpha_k}; \quad \alpha_k = \frac{\eta}{z_{rk}}.
\end{equation}

As has been pointed out elsewhere \cite{esarey,salamin92a,salamin92b}, $f_k$ is a function that must be adopted to appropriately represent the wavenumber distribution in $k$-space of the initial pulse, and whose Fourier transform will be related below to the spatio-temporal pulse envelope. Unfortunately, the Fourier transform of $a_k$ according to Eq. (\ref{ak}) cannot be evaluated analytically, in general. Resort to approximation is, therefore, inevitable. Viewing $\psi_k$ as a function of $k'\equiv k+k_0$, the following series expansion may be the natural approach to follow, namely
\begin{eqnarray}\label{psikexpansion}
  \psi_k &=& \sum_{m=0}^\infty\frac{(k'-k_0)^m}{m!} \left.\frac{\partial^m\psi_k}{\partial k'^m}\right|_{k'=k_0},\nonumber\\
         &=& \sum_{m=0}^\infty \frac{k^m}{m!} \psi_0^{(m)};\quad \psi_0^{(m)} \equiv\left.\frac{\partial^m\psi_k}{\partial k^m}\right|_{k=0}.
\end{eqnarray}
With the help of this expansion, the vector potential amplitude may be written as
\begin{equation}\label{aa}
  a(\rho,\eta,\zeta) = \sum_{m=0}^\infty \frac{\psi_0^{(m)}}{m!} F_m(\zeta),
\end{equation}
where
\begin{equation}\label{Fm}
  F_m(\zeta) = \frac{1}{\sqrt{2\pi}}\int_{-\infty}^\infty \left(f_k k^m\right) e^{ik\zeta}dk,
\end{equation}
is Fourier transform of the product $f_k\psi_k$. In anticipation of the conclusion, to be arrived at shortly, that terms in the sum (\ref{aa}) beyond the first will contribute negligibly in real applications, Eq. (\ref{aa}) will be replaced by the truncated series
\begin{equation}\label{atruncate}
  a^{(n)}(\rho,\eta,\zeta) \approx \sum_{m=0}^n \frac{\psi_0^{(m)}}{m!} F_m(\zeta).
\end{equation}

At this stage, the complex amplitude will be written as $A_0=a_0e^{i\varphi_0}$, in which $\varphi_0$ is an initial phase and $a_0$ is a real amplitude for the exact vector potential. The status of $a_0$ will be slightly modified in the approximate solutions to be presented below, by introducing model-dependent normalization factors, defined appropriately at each level of truncation. With this, the truncated vector potential (to order of truncation $n$) takes the form
\begin{equation}\label{Atruncate}
 A^{(n)}(\rho,\eta,\zeta) \approx a_0 e^{i\varphi_0+ik_0\zeta}\sum_{m=0}^n \frac{\psi_0^{(m)}}{m!} F_m(\zeta)
\end{equation}
Equation (\ref{Atruncate}) can be employed to obtain the electric and magnetic fields, to any desired order of truncation. However, on account of the conclusion to be made shortly for some initial frequency spectra, the zeroth-order terms may be the most dominant ones and only the lowest-order corrections may be necessary. For book-keeping purposes, explicit expressions for $\psi_0^{(m)}$, with $m = 0-3$, are \cite{salamin92a,salamin92b}
\begin{equation}
\label{ps00}  \psi_0^{(0)} = \beta e^{-\beta\rho^2},
\end{equation}
\begin{equation}
\label{ps01}  \psi_0^{(1)} = \frac{i\alpha}{k_0}(1-\beta\rho^2) \beta^2  e^{-\beta\rho^2},
\end{equation}
\begin{equation}
\label{ps02}  \psi_0^{(2)} = \frac{i\alpha}{k_0^2}\left[-2+(4\beta-2)\rho^2+i\alpha\beta^2\rho^4\right] \beta^3 e^{-\beta\rho^2},
\end{equation}
\begin{eqnarray}
\label{ps03}  \psi_0^{(3)} &=& \frac{i\alpha}{k_0^3}\left[6+6(2-3\beta)\rho^2+3i\alpha\beta(1-3\beta)\rho^4 \right.\nonumber\\
               & &\left. +\alpha^2\beta^3\rho^6\right] \beta^4 e^{-\beta\rho^2}.
\end{eqnarray}
In Eqs. (\ref{ps00})-(\ref{ps03})
\begin{equation}\label{zr}
  \beta = \frac{1}{1+i\alpha};\quad \alpha = \frac{\eta}{z_r};\quad z_r \equiv z_{r0} = \frac{1}{2}k_0w_0^2,
\end{equation}
where $z_r$ is the Rayleigh length.
The zeroth-order electric and magnetic fields of an ultrashort and tightly focused laser pulse will be derived from the above equations, fully analytically. The first-order, and possibly higher-order corrections, can in principle be used in numerical calculations.

\section{The fields}\label{sec:EandB}

The radially polarized $\bm{E}$ and $\bm{B}$ fields will be obtained below from the one-component axially polarized vector potential \begin{equation}\label{A}
  \bm{A} = \hat{z} A;\quad A = a_0 a e^{i\varphi_0+ik_0\zeta},
\end{equation}
where $\hat{z}$ is a unit vector in the direction of propagation of the pulse, taken along the $z$-axis of a cylindrical coordinate system, together with the associated scalar potential \cite{mcdonald,salaminNJP8,salamin92a,salamin92b}. Employing SI units, the radial and axial electric field components may be obtained, respectively, from \cite{salamin92a}
\begin{equation}
\label{Erform} E_r = -\frac{c^2}{R} \frac{\partial}{\partial r} \left(\frac{\partial A}{\partial z}\right) - \frac{c^2}{R^2}
    \left(\frac{\partial A}{\partial z}\right) \frac{\partial}{\partial r} \left(\frac{1}{a} \frac{\partial a}{\partial t}\right),
\end{equation}
\begin{equation}
\label{Ezform} E_z = -\frac{\partial A}{\partial t}- \frac{c^2}{R} \frac{\partial^2 A}{\partial z^2} - \frac{c^2}{R^2}
    \left(\frac{\partial A}{\partial z}\right) \frac{\partial}{\partial z} \left(\frac{1}{a} \frac{\partial a}{\partial t}\right),
\end{equation}
in which
\begin{equation}\label{Rform}
   R = ick_0-\frac{1}{a}\frac{\partial a}{\partial t}.
\end{equation}
Furthermore, the only (azimuthal) magnetic field component will follow from
\begin{equation}\label{Bthetaform}
  B_\theta = -\frac{\partial A}{\partial r}.
\end{equation}

\begin{figure}
\includegraphics[width=8cm]{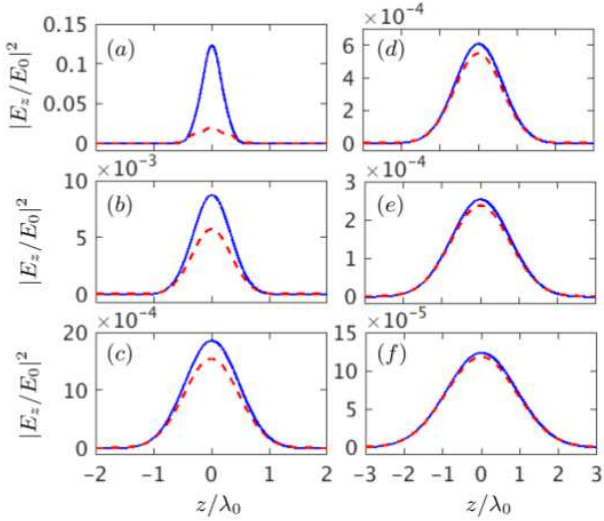}
\caption{(Color online) The initial ($t = 0$) scaled intensity $|E_z/E_0|^2$ along the propagation direction ($\rho = 0$) of a radially polarized pulse prepared from a Gaussian initial spectrum and using the zeroth-order field component $E_z$ (blue solid lines) and when the first-order correction to $E_z$ is included (red dashed lines). Values of $L = w_0$ increase from (a) $0.5\lambda_0$ to (f) $3\lambda_0$ in steps of $0.5\lambda_0$. For all graphs, $\varphi_0 = 0$. }
\label{fig1}
\end{figure}

Derivation of analytic expressions for the electric and magnetic fields from the appropriate vector potential will be done below. The starting point along this path is a choice that must be made for an appropriately defined initial pulse spectrum. In the next two subsections, two such choices will be made.

\subsection{An initial Gaussian spectrum}

An initial Gaussian spectrum is considered first, following Esarey {\it et al.} \cite{esarey} and our own earlier work \cite{josab33}. Employing the same notation as in \cite{esarey,josab33} the initial wavenumber distribution will be given by
\begin{equation}\label{fkgauss}
	f_k = \frac{\sigma}{k_0}\left(1+\frac{k}{k_0}\right) \exp\left(-\frac{k^2\sigma^2}{2k_0^2}\right),
\end{equation}
in which $\sigma$ is the pulse's initial full-width-at-half-maximum, given in terms of its spatial extension in the forward direction, $L\sim c\tau_0$, where $\tau_0$ is the temporal pulse duration, by $\sigma = k_0L/(2\sqrt{2\ln 2})$.

\begin{figure}[t]
\includegraphics[width=8cm]{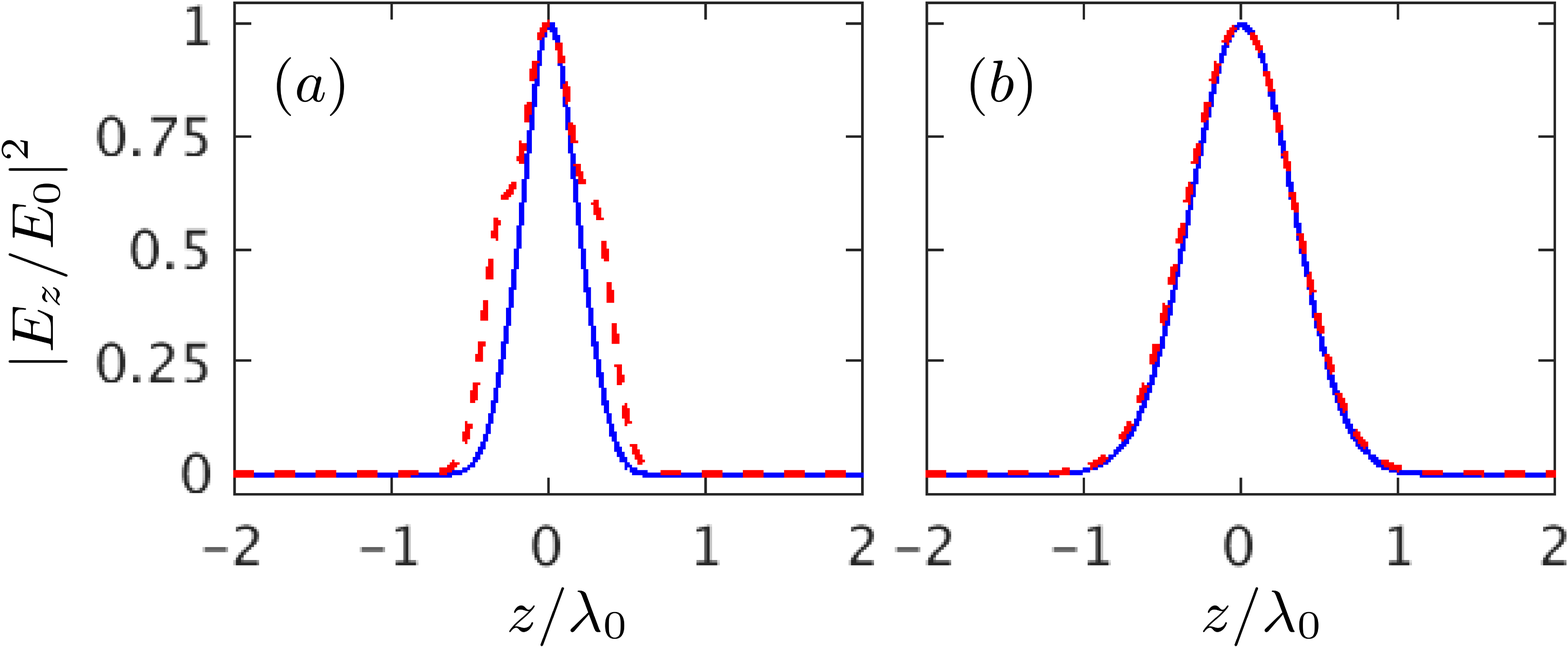}
\caption{(Color online) Same as Fig. \ref{fig1}a and \ref{fig1}b, but using the normalization factors $N_G^{(0)}$ and $N_G^{(1)}$. The blue solid lines are for the zeroth-order field component, while the red dashed lines are based on $E_z$ up to, and including, first-order.}
\label{fig2}
\end{figure}

To move on to finding the vector potential amplitude according to Eq. (\ref{atruncate}) one needs to evaluate the integral (\ref{Fm}) first. To any order of truncation $m$, one has
\begin{eqnarray}\label{FmG}
  F^G_m(\zeta) &=& \sqrt{\frac{2^m}{4\pi}}\left(\frac{k_0}{\sigma}\right)^m \left\{\frac{\sqrt{2}}{\sigma}\left[(-1)^{m+1}+1\right] \right.\nonumber\\
  & &\left.\times \Gamma \left(\frac{m+2}{2}\right)\left[ ~_1F_1\left(\frac{m+2}{2},\frac{1}{2},-\frac{k_0^2\zeta^2}{2\sigma^2}\right) \right.\right.\nonumber\\
  & &\left.\left.+\left(ik_0\zeta\right) ~_1F_1\left(\frac{m+2}{2}, \frac{3}{2}, -\frac{k_0^2 \zeta^2}{2 \sigma^2}\right)\right]\right.\nonumber\\
  & &\left.+\left[(-1)^m+1\right] \Gamma \left(\frac{m+1}{2}\right)\right.\nonumber\\
  & &\left.\times \left[ ~_1F_1\left(\frac{m+1}{2},\frac{1}{2},-\frac{k_0^2\zeta^2}{2 \sigma^2}\right)+\left(\frac{ik_0\zeta}{\sigma^2}\right)\right.\right.\nonumber\\
  & &\left.\left.\times(m+1) ~_1F_1\left(\frac{m+3}{2},\frac{3}{2},-\frac{k_0^2\zeta^2}{2\sigma ^2}\right)\right]\right\}.
\end{eqnarray}
In this equation $_1F_1$ stands for a confluent hypergeometric function, and the superscript $G$ stands for {\it Gaussian}. To lowest-order of truncation, and according to Eq. (\ref{Fm}) or (\ref{FmG})
\begin{equation}\label{F0G}
	F^G_0(\zeta) = q \exp\left(-\frac{k_0^2\zeta^2}{2\sigma^2}\right);
    \quad q = 1+\frac{ik_0\zeta}{\sigma^2},
\end{equation}
will give, at $t=0$, the initial spatial pulse envelope. To the same order, Eqs. (\ref{atruncate}) and (\ref{ps00}) yield
\begin{equation}\label{aG}
  a^{(0)} = \beta e^{-\beta\rho^2} F^G_0(\zeta).
\end{equation}
After some elaborate algebra, Eqs. (\ref{Erform})-(\ref{Bthetaform}) give
\begin{equation}
\label{Er0}  E_r^{(0)} = E_G\left(\frac{\rho\beta^2}{k_0w_0}\right)G_r^{(0)},
\end{equation}
\begin{equation}
\label{Ez0} E_z^{(0)} = E_G\left(\frac{i\beta}{2k_0z_r}\right)
        G_z^{(0)},
\end{equation}
\begin{equation}
\label{Btheta0} cB_{\theta}^{(0)} = E_G\left(\frac{2\rho\beta^2 }{k_0w_0}\right) q.
\end{equation}
In Eqs. (\ref{Er0})-(\ref{Btheta0})
\begin{equation}\label{EG}
   E_G = N_G^{(0)}E_0\exp\left(i\varphi_0+ik_0\zeta -\frac{k_0^2\zeta^2}{2\sigma^2} -\beta\rho^2\right),
\end{equation}
is common to all three field components. To maintain the general structure of the defining equations of the electric field components, namely, Eqs. (\ref{Erform}) and (\ref{Ezform}) the following functions are introduced
\begin{equation}\label{Gr}
 G_r^{(0)} = \frac{ic}{z_r}\left(\frac{G_1}{R_G}-\frac{ic\beta G_2}{2z_rR_G^2}\right),
\end{equation}
and
\begin{equation}\label{Gz}
  G_z^{(0)}=\frac{2qz_r}{ic} R_G-\frac{ic}{2z_r}\left(\frac{G_3}{R_G}-\frac{ic G_4}{2z_rR_G^2}\right).
\end{equation}

\begin{figure}
\includegraphics[width=8cm]{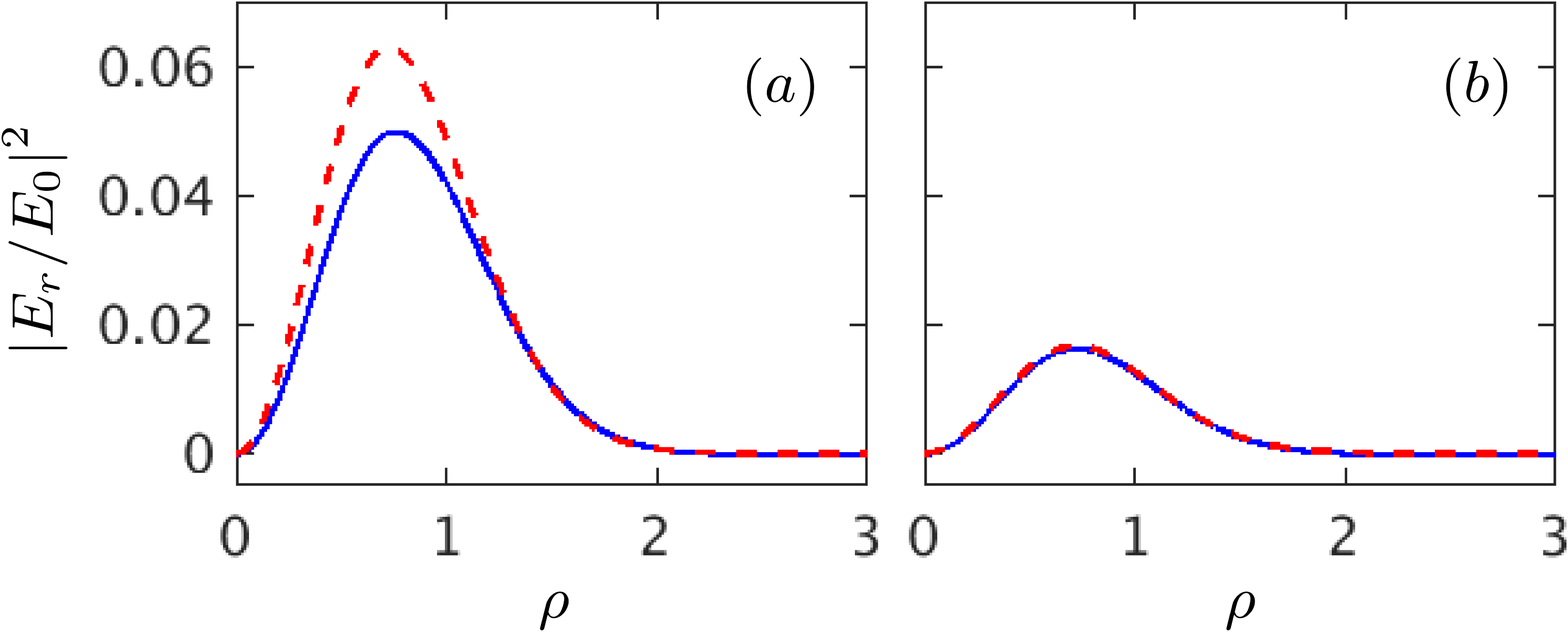}
\caption{(Color online) Same as Fig. \ref{fig1}, but for the scaled intensity $|E_r/E_0|^2$ in the focal plane ($z = 0$). Values of the parameters $w_0 = L$ are: (a) $0.5\lambda_0$, and (b) $\lambda_0$.}
\label{fig3}
\end{figure}

With $E_0 = ck_0a_0$ and $N_G^{(0)}$ given by Eq. (\ref{N0G}) below, the remaining quantities in Eqs. (\ref{Gr}) and (\ref{Gz}) are
\begin{equation}\label{RG}
   R_G = \left(\frac{ic}{2qz_r}\right)\left[q\beta(1-\beta\rho^2)+2k_0z_r\left(q^2 + \frac{1}{\sigma^2}\right)\right],
\end{equation}
which follows from Eq. (\ref{Rform}) using the Gaussian spectrum, and
\begin{equation}\label{G1}
   G_1 = -q\beta+G_2,
\end{equation}
\begin{equation}\label{G2}
   G_2 = q\beta(-1+\beta\rho^2)+2k_0z_r\left[q^2 + \frac{1}{\sigma^2}\right],
\end{equation}
\begin{eqnarray}\label{G3}
G_3 &=& q\beta^2(2-4\beta\rho^2 +\beta^2 \rho^4)\nonumber\\
    & &+4k_0z_r\beta
   \left(-1+\beta\rho^2\right)\left[q^2+\frac{1}{\sigma^2}\right]\nonumber\\
    & &+4k_0^2z_r^2q \left[q^2+\frac{3}{\sigma^2}\right],
\end{eqnarray}
\begin{eqnarray}\label{G4}
G_4 &=& G_2\left[\beta^2 (-1+2 \beta \rho^2)\right. \nonumber\\
    & &\left.+
 \left(\frac{2k_0z_r}{q\sigma}\right)^2 \left(q^2-\frac{1}{\sigma^2}\right)\right].
\end{eqnarray}

Note that the radial electric field and azimuthal magnetic field components vanish identically at all points on the propagation axis, where $\rho=0$. For applications in which corrections to the field components beyond first-order need not be considered, a {\it normalization factor}, $N^{(0)}$, will be introduced, whose value is to be found by requiring that $E_z(0,0,0) = E_0$. Dropping correction terms of first-order of truncation and beyond, where applicable, it is suggested here to let $a_0 \to a_0N_G^{(0)}$, and to use this requirement to obtain
\begin{equation}\label{N0G}
  N_G^{(0)} = -\frac{2i\left[\sigma^2/2+k_0z_r \left(\sigma ^2+1\right)\right]^2}{\sigma^4+\sigma^2+2 k_0z_r \left(2 \sigma^4+3\sigma^2+3\right)},
\end{equation}
in order to partially compensate for the loss of accuracy.
\begin{figure}[t]
\includegraphics[width=8cm]{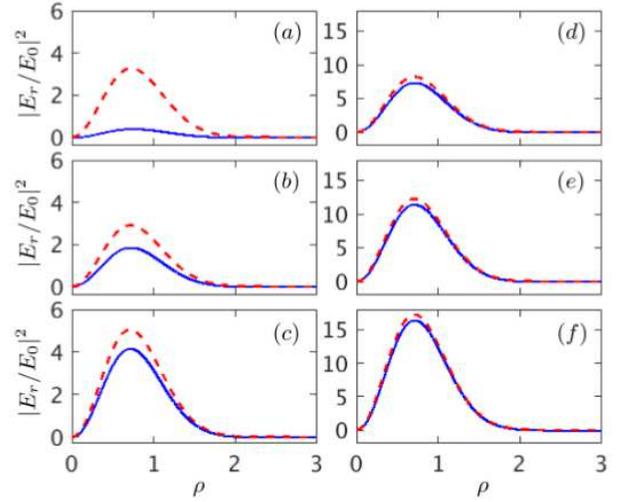}
\caption{(Color online) Same as Fig. \ref{fig2}, but for the scaled intensity $|E_r/E_0|^2$ in the focal plane ($z = 0$). Values of the parameters $w_0 = L$ range from (a) $0.5\lambda_0$ to (f) $3\lambda_0$, in steps of $0.5\lambda_0$. As in the other figures, solid blue: zeroth-order, and red dashed: including the first-order corrections.}
\label{fig4}
\end{figure}

For applications which require inclusion of the first-order corrections to the fields given by Eqs. (\ref{Er0})-(\ref{Btheta0}) the quantity
\begin{equation}\label{F1G}
  F^G_1(\zeta) = \frac{k_0}{\sigma^2} \left[1+ik_0\zeta-\left(\frac{k_0\zeta}{\sigma}\right)^2\right] \exp\left[-\frac{k_0^2\zeta^2}{2\sigma^2}\right],
\end{equation}
is needed in addition to $F_0$. To that order, $a_0 \to a_0 N_G^{(1)}$, in which the corresponding normalization factor is
\begin{equation}\label{N1G}
	N_G^{(1)} = -\frac{i\sigma^2 \left[-1+\sigma^2+2k_0 z_r(1+\sigma^2)\right]^2}{2 \left[2+3 \sigma^2-2
   \sigma^4+\sigma^6-4k_0z_r(1-\sigma^6)\right]}.
\end{equation}

To assess the relative importance of the first-order corrections to the zeroth-order field components, Fig. \ref{fig1} shows the scaled intensity, $|E_z/E_0|^2$, along the propagation direction, without the normalization factors, of a pulse, with and without these corrections. It is clear from the figure that the zeroth-order expression is sufficient for describing the field component $E_z$, so long as the values of $w_0 = L>\lambda_0$. For $w_0 = L = 0.5\lambda_0$, adding the first-order correction term changes the Gaussian shape of the pulse quite visibly, indicating that the representation may not be valid down to that level of tight focusing.

\begin{figure}
\includegraphics[width=8cm]{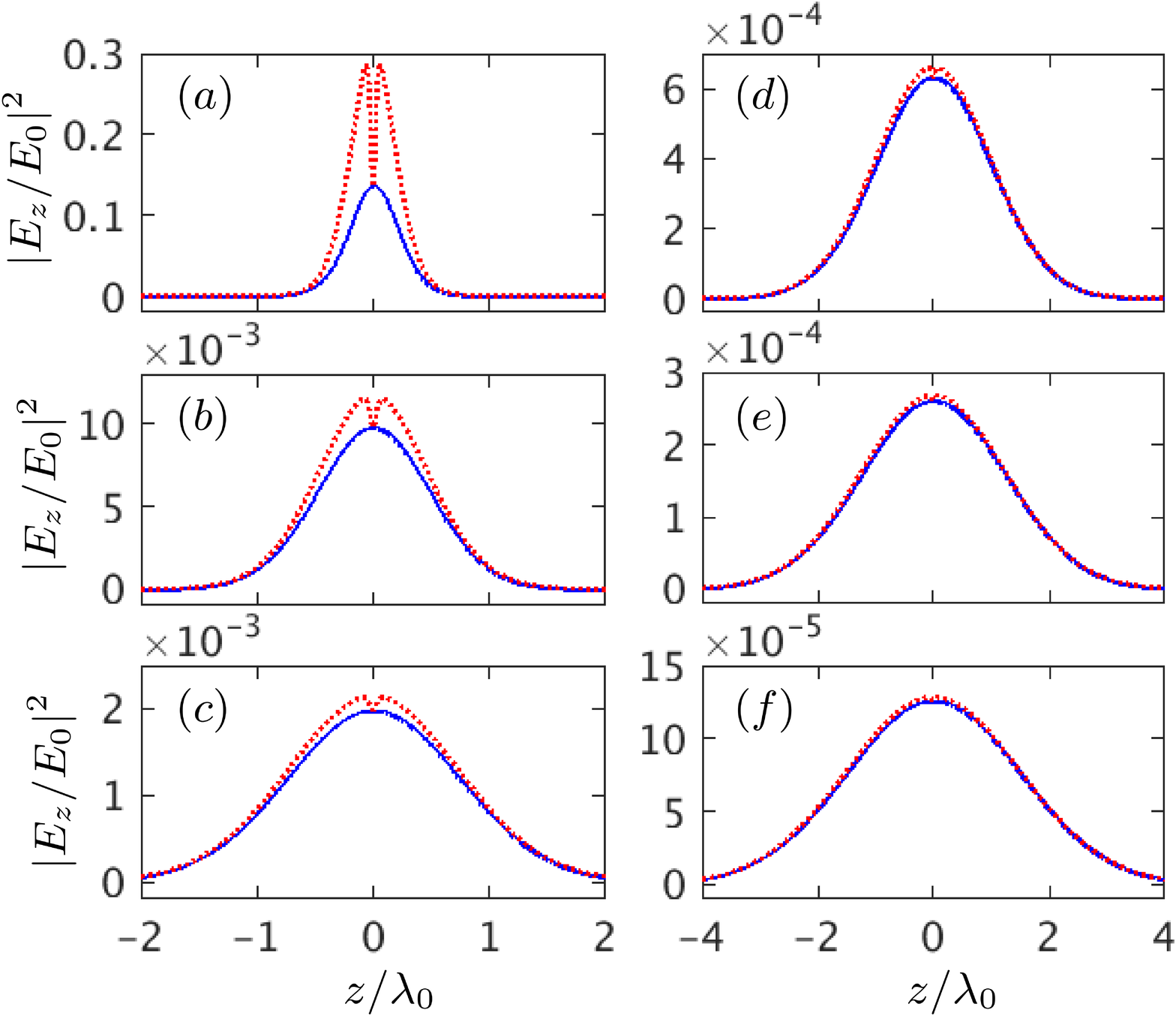}
\caption{(Color online) Initial ($t=0$) intensity profiles $|E_z/E_0|^2$ vs. distance along the propagation direction, for a radially polarized laser pulse evolving from a Poisson initial spectrum. The solid blue lines are for a description based on the zeroth-order field component $E_z$ and the red dotted lines are for $E_z$ up to second-order truncation. From (a) to (f) the values of $w_0 = L$ vary from $0.5\lambda_0$ to $3\lambda_0$ in steps of $0.5\lambda_0$, and the corresponding parameter $s$ = 4.37, 19.226, 43.91, 78.45, 122.87, 177.15, respectively, according to Eq.~(\ref{s}).}
\label{fig5}
\end{figure}

To make up for the compromised accuracy, which results from dropping the higher-order corrections, a normalization factor has been employed in the fashion described above. In Fig. \ref{fig2}, the same intensity profile, $|E_z/E_0|^2$, is shown for the cases (a) and (b) of Fig. \ref{fig1}, but with the normalization factors included. Inclusion of the first-order terms does not result in any visible corrections to the intensity profiles for all values of $w_0 = L \ge\lambda_0$, and thus those terms can safely be dropped when the normalization factors are employed. Unfortunately, the range of validity of the zeroth-order fields cannot clearly be extended down to a pulse of sub-wavelength waist radius $w_0$ and/or axial length $L$. Neither is the Gaussian shape restored fully under these conditions.

Next, the intensity profile $|E_r/E_0|^2$, derived from the radial component of the electric field, is investigated in the initial focal plane ($t = 0 = z$). Examples are shown in Figs. \ref{fig3} and \ref{fig4}. It is obvious from Fig. \ref{fig3}, and for the parameter set used, that addition of the first-order correction alters the profile visibly only for $w_0 = L = 0.5\lambda_0$, as shown in Fig. \ref{fig3}a. For $w_0 = L \ge\lambda_0$, on the other hand, the zeroth-order description seems to be sufficient, and the first-order corrections may safely be dropped.

\subsection{An initial Poisson spectrum}

\begin{figure}
\includegraphics[width=8cm]{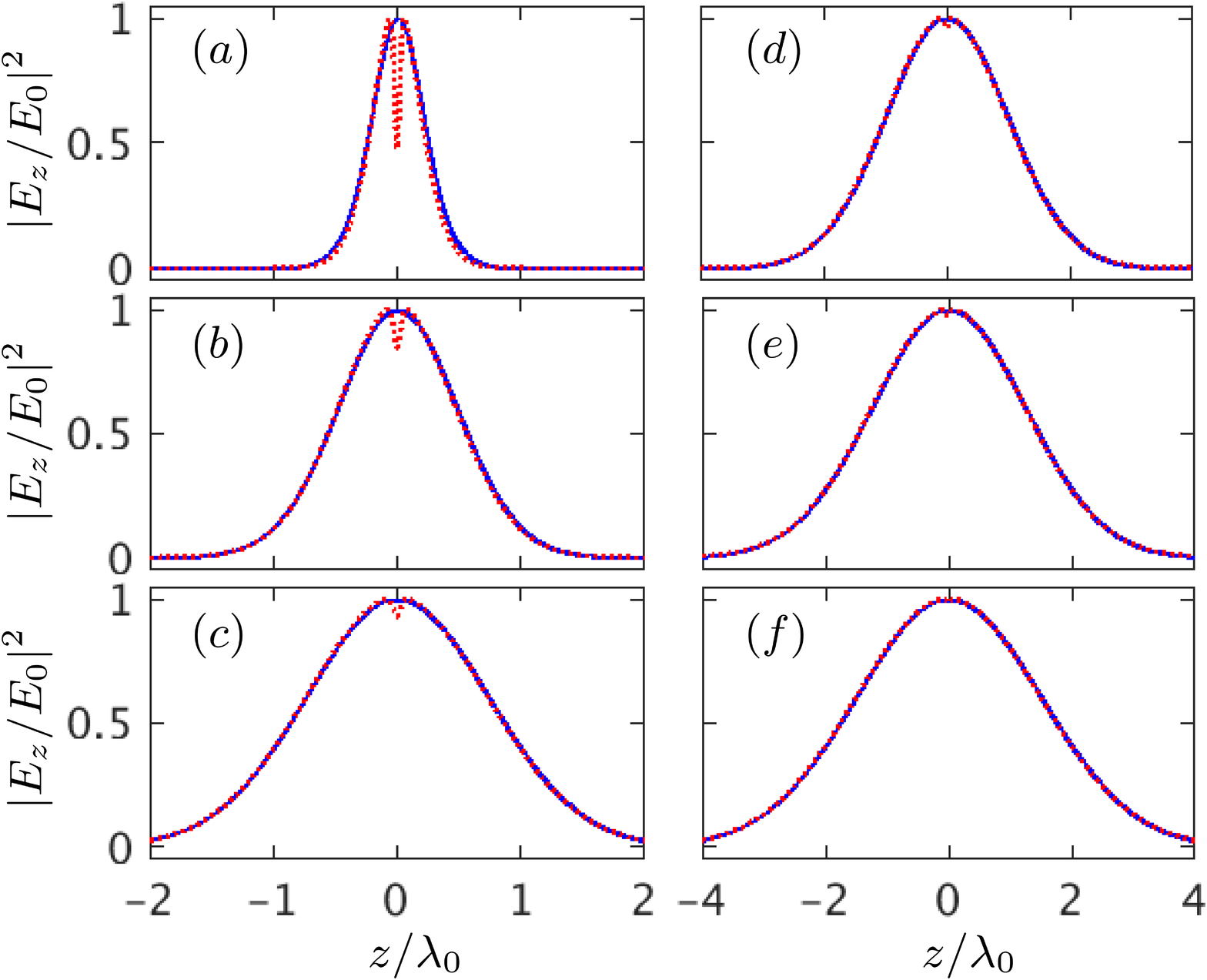}
\caption{(Color online) Same as Fig. \ref{fig5}, but using the normalization factors specific to a radially polarized pulse which evolves from an initial Poisson spectrum.}
\label{fig6}
\end{figure}

As has already been pointed out elsewhere \cite{esarey,josab33} the Gaussian spectrum (\ref{fkgauss}) has, inherent in it, negative frequencies. The first term in Eq. (\ref{fkgauss}) has been introduced specifically \cite{esarey} to eliminate the frequency corresponding to $k=-k_0$. An initial spectrum which involves only physical frequencies, to be employed next, is the Poissonian \cite{april1,april2,varin1}
\begin{eqnarray}\label{poissonk}
    f_k = \sqrt{2\pi}\left(\frac{s}{k_0}\right)^{s+1}\frac{k^s e^{-s k/k_0}}{\Gamma(s+1)}H(k),
\end{eqnarray}
where $s$ is a real positive parameter, $\Gamma(x)$ is the Gamma function, and $H(k)$ is a unit step function, introduced precisely to exclude the (unphysical) negative frequencies. The parameter $s$ is related to the laser pulse duration $\tau_0$, or root-mean-square of the temporal distribution, through \cite{april1,april2}
\begin{eqnarray}\label{s}
    \tau_0=\frac{2s}{\omega_0 \sqrt{2s-1}}.
\end{eqnarray}
Equation (\ref{s}) is quadratic in $s$. In subsequent developments and the examples to be investigated and presented graphically, values of $s>1$ will be considered only.

Fourier transform of $f_k$ for the Poisson spectrum, for an arbitrary order of truncation $m$, follows immediately from Eq. (\ref{Fm}). The result is
\begin{equation}\label{FmP}
  F^P_m(\zeta) = \left(\frac{k_0}{s}\right)^m\frac{\Gamma (m+s+1)}{\Gamma(s+1)} h^{-m-s-1},
\end{equation}
in which the superscript $P$ stands for {\it Poisson}, and
\begin{equation}\label{h}
  h=1-\frac{ik_0\zeta}{s}.
\end{equation}
Next, the lowest-order fields will be derived from the vector potential, following the procedure outlined in Sec. \ref{sec:back} above. The main equations for the zeroth-order fields will be spelled out explicitly here, together with the defining equations for the quantities of relevance to them.

\begin{figure}
\includegraphics[width=8cm]{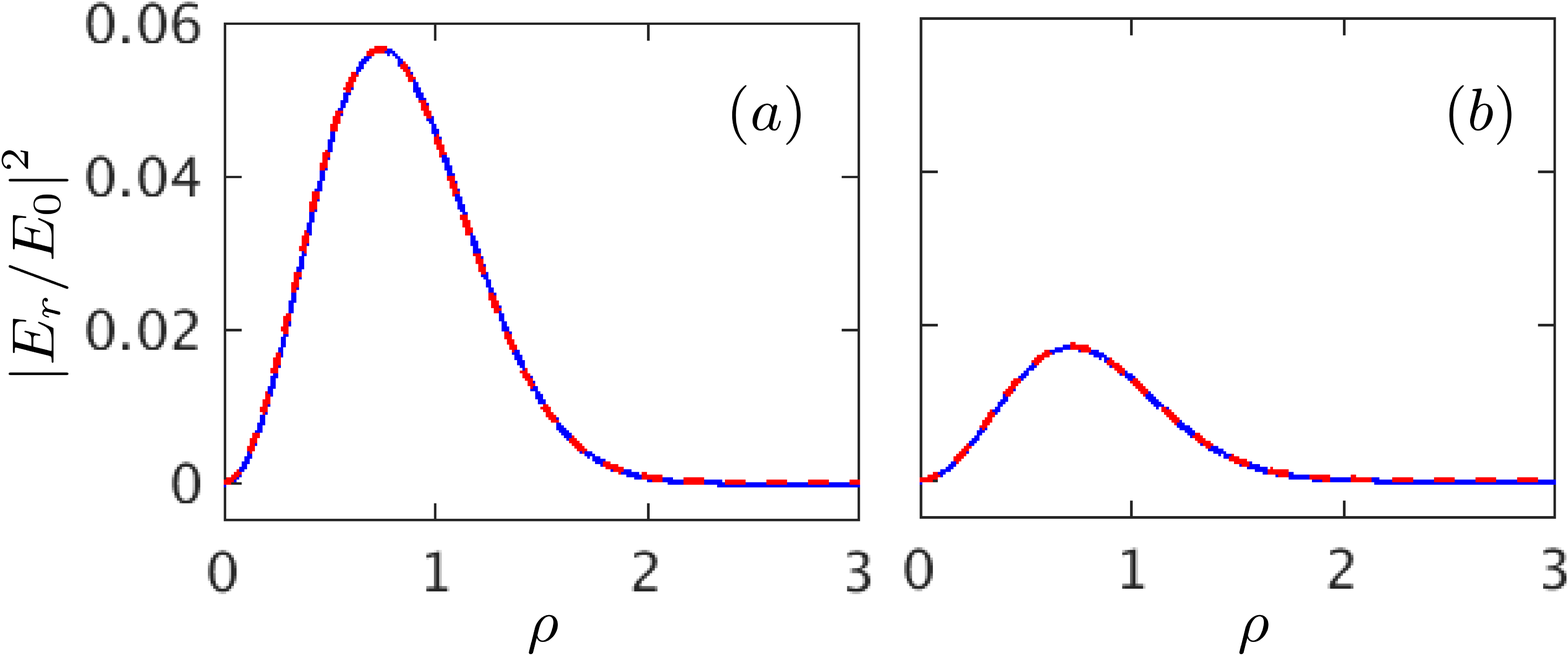}
\caption{(Color online) Same as Fig. \ref{fig5}, but for the scaled intensity $|E_r/E_0|^2$ in the focal plane ($z = 0$) of the pulse which evolves from an initial Poisson spectrum. Values of the parameters $w_0 = L$ are: (a) $0.5\lambda_0$, and (b) $\lambda_0$, and $s$ = 4.37, 19.226, respectively.}
\label{fig7}
\end{figure}

The starting point is a vector potential similar to that given by Eq. (\ref{A}) with an expression for $a(\rho,\eta,\zeta)$ exactly the same as in Eq. (\ref{aG}) apart from
\begin{equation}\label{F0P}
F^P_0(\zeta)=h^{-s-1},
\end{equation}
the Fourier transform of $f_k$. The following field components follow from Eqs. (\ref{Erform})-(\ref{Bthetaform})
\begin{equation}\label{Er0P}
E_r^{(0)} =  E_P\left(\frac{\rho\beta^2 }{k_0 w_0}\right) P_r^{(0)},
\end{equation}
\begin{equation}
\label{Ez0P} E_z^{(0)} = E_P\left(\frac{i\beta}{2k_0z_r}\right)
        P_z^{(0)},
\end{equation}
\begin{equation}\label{Btheta0P}
cB_{\theta}^{(0)} = E_P\left(\frac{2\rho\beta^2}{k_0w_0}\right)h.
\end{equation}
Here, too, $E_0 = ck_0a_0$, and
corresponding to Eq. (\ref{EG}) of the Gaussian distribution, one has for the Poisson spectrum
\begin{equation}\label{EP}
   E_P = N_P^{(0)}E_0e^{i\varphi_0+ik_0\zeta-\beta\rho^2} \left[1-\frac{ik_0\zeta}{s}\right]^{-s-2}.
\end{equation}
The structure of the defining equations (\ref{Erform}) and (\ref{Ezform}) is maintained here, too, which requires introduction of the Posisson spectrum functions
\begin{equation}\label{Pr}
 P_r^{(0)} = \frac{ic}{z_r}\left(\frac{P_1}{R_P}-\frac{ic\beta P_2}{2z_rR_P^2}\right),
\end{equation}
and
\begin{equation}\label{Pz}
  P_z^{(0)}=\frac{2hz_r}{ic} R_P-\frac{ic}{2z_r}\left(\frac{P_3}{R_P}-\frac{ic P_4}{2z_rR_P^2}\right).
\end{equation}
In Eqs. (\ref{Pr}) and (\ref{Pz})
\begin{equation}\label{RP}
   R_P = \left(\frac{ic}{2hz_r}\right)\left[h\beta(1-\beta\rho^2) +2k_0z_r\left(1+\frac{1}{s}+h\right)\right],
\end{equation}
is obtained from (\ref{Rform}) employing parameters of the Poisson spectrum, and
\begin{equation}\label{P1}
   P_1 = -h\beta+P_2,\\
\end{equation}
\begin{equation}\label{P2}
   P_2 = h\beta(-1+\beta\rho^2)+2k_0z_r\left(1+\frac{1}{s}+h\right),\\
\end{equation}
\begin{eqnarray}\label{P3}
P_3 &=& h\beta^2(2-4\beta\rho^2 +\beta^2 \rho^4)\nonumber\\
    & &+4k_0z_r\beta
   \left(-1+\beta\rho^2\right)\left[1+\frac{1}{s}+h\right]\nonumber\\
    & &+4k_0^2z_r^2 \left[\frac{2}{h s^2}+\frac{3}{h s}+\frac{1}{h}+\frac{2}{s}+h+2\right],
\end{eqnarray}

\begin{equation}\label{P4}
P_4 = P_2\left[\beta^2 (-1+2 \beta \rho^2) +
 \left(\frac{2 k_0 z_r}{h s}\right)^2 \left(s+1\right)\right].
\end{equation}
Furthermore, the following is a normalization factor similar to (\ref{N0G}) of the Gaussian distribution
\begin{equation}\label{N0P}
N_P^{(0)}=\frac{-i\left[s+2k_0z_r\left(1+2s\right)\right]^2}{2s\left(1+2s\right) +4k_0z_r\left(1+7s+8s^2\right)}.
\end{equation}

\begin{figure}
\includegraphics[width=8cm]{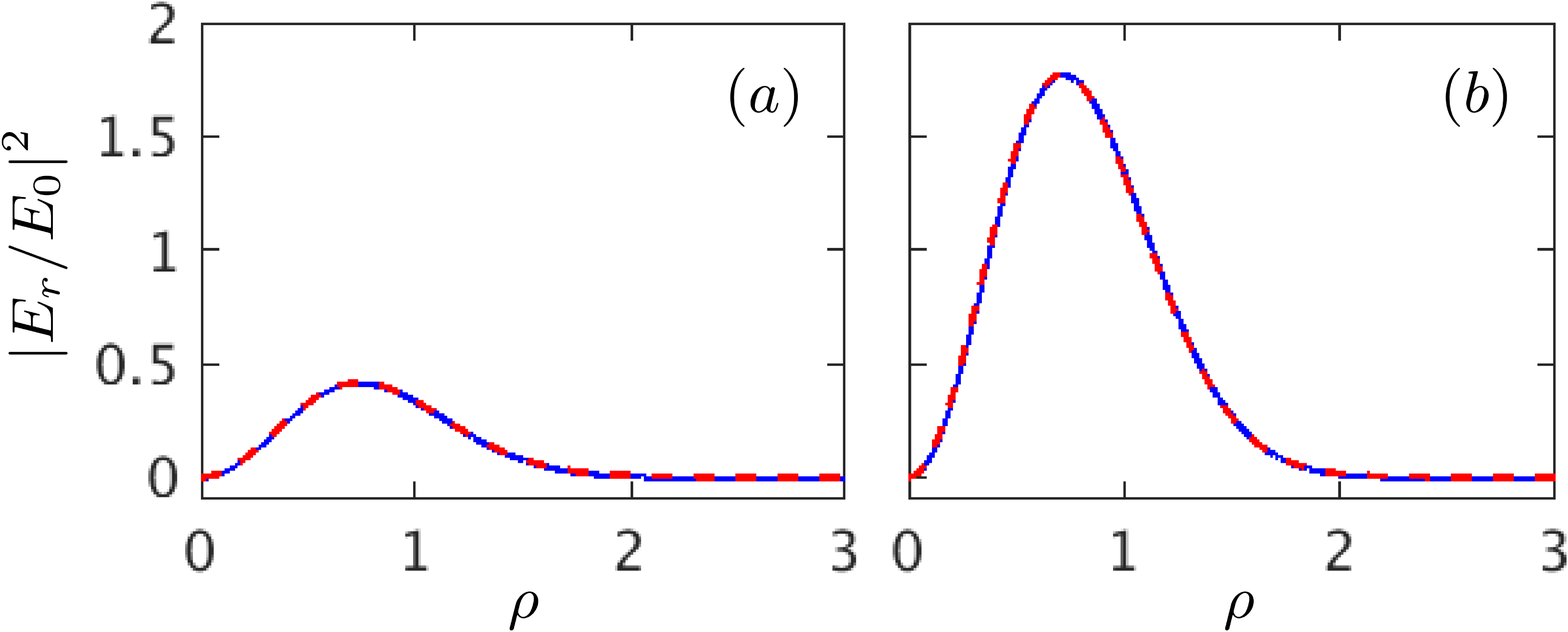}
\caption{(Color online) Same as Fig. \ref{fig6}, but using the normalization factors specific to a radially polarized pulse which evolves from an initial Poisson spectrum.}
\label{fig8}
\end{figure}

To save space in this paper, the first-order corrections to the fields obtained in this section will not be given. Only the corresponding pulse-shape function $F^P_1$ and the normalization factor $N_P^{(1)}$ will be given, respectively, by
\begin{equation}\label{F1P}
  F^P_1(\zeta) = k_0\left(\frac{s+1}{s}\right) h^{-s-2},
\end{equation}
and
\begin{equation}\label{NP1}
  N_P^{(1)} = \frac{-is\left[1-2k_0z_r\left(1+2s\right)\right]^2}{2\left[1+11s+10s^2+4s^3 -4k_0z_r\left(1+5s+5s^2\right)\right]}.
\end{equation}

The set of figures 5-8 correspond, one-to-one and respectively, to the set 1-4, albeit for the Poisson initial frequency spectrum. One major distinction between the two sets should be pointed out, though. In Figs. 1-4 the field components employed are described by zero- and first-order truncation terms, whereas Figs. 5-8 have been calculated using up to and including second-order truncation corrections. Note that, in Figs. \ref{fig5} and \ref{fig6}, adding further correction terms does not result in good convergence of the displayed curves, for $w_0 = L\le\lambda_0$. Upon further scrutiny, this behaviour has been found to be due to the odd-order truncation terms. By contrast, in the radial intensity profiles displayed in Figs. \ref{fig7} and \ref{fig8}, convergence occurs already at the first-order correction and, therefore, terms beyond zeroth-order, in the expressions which model the radial electric field component $E_r$, can safely be dropped. An accurate description of $E_r$ of an ultra-short and tightly-focused laser pulse may adequately be described by the lowest order term, down to the range $w_0 = L = \lambda_0/2$.

\section{Summary and Conclusions}\label{sec:conc}

This paper has been devoted to the analytic calculation of the electric and magnetic fields, to lowest-order truncation, of an ultra-short and tightly-focused laser pulse, of the radially polarized variety, propagating in a vacuum, from vector and scalar potentials linked by the Lorentz condition. Two initial wavenumber distributions have been used to synthesize the fields of the pulse by proper Fourier transformation. Every effort has been exerted to develop and present the final results from both distributions, similarly. Similar quantities have been introduced in the descriptions of the results, distinguished from one another merely by subscripts $G$ (for Gaussian) and $P$ (for Poisson) to facilitate comparison and contrast. Whereas explicit expressions have been reported only for the most dominant (zeroth-order) terms, a complete program has also been outlined, including some important analytic steps of basic relevance,  for calculating the field components numerically, to any desired order of truncation. Scaled initial intensity profiles $|E_z/E_0|^2$ and $|E_r/E_0|^2$, based on our analytic (zeroth-order) and numerical (first- and second-order) have been displayed graphically, employing a set of parameters which emphasizes the ultra-short and tightly-focused characteristics of the pulse.

Our main conclusions from this work may be summarized as follows. The analytic expressions reported above for the zeroth-order components of the fields are compact, and may be found quite handy for many applications. Terms of higher order than the zeroth, however, turned out to be quite cumbersome and have, thus, been left out. Alternatively, the program of calculating the fields numerically can be resorted to, if needed. It has been demonstrated, by showing initial intensity distributions, that the zeroth-order expressions are sufficient for describing all of the field components when the Gaussian spectrum is employed, and for pulse axial length, $L$, and waist radius at focus, $w_0$, down to $\lambda_0/2$. For the Poisson spectrum, the same conclusion has been arrived at for only the radial component $E_r$. More (higher-order) terms are needed to adequately model the axial component $E_z$, for the same parameter set just mentioned.

\begin{acknowledgments}

YIS is supported by a Faculty Research Grant (FRG-13) from the American University of Sharjah.

\end{acknowledgments}


\begin{thebibliography}{100}

\bibitem{eli} ELI: {\it Extreme Light Infrastructure}, \url{http://www.eli-beams.eu/}.
\bibitem{hiper} HiPER: {\it The High Power laser Energy Research (HiPER) facility}, \url{http://www.hiper-laser.org}.
\bibitem{xfel} XFEL: {\it The European X-Ray Laser Project XFEL}, \url{http://xfel.eu/}.

\bibitem{marklund} M. Marklund and P. K. Shukla, Rev. Mod. Phys. {\bf 78}, 591 (2006).
\bibitem{mourou} G. A. Mourou, T. Tajima, and S. V. Bulanov, Rev. Mod. Phys.
{\bf 78}, 309 (2006).
\bibitem{esarey2009} E. Esarey, C. B. Schroeder, and W. P. Leemans, Rev. Mod. Phys. {\bf 81}, 1229 (2009).
\bibitem{piazza} A. Di Piazza, C. M\"uller, K. Z. Hatsagortsyan, and C. H. Keitel, Rev. Mod. Phys. {\bf 84}, 1177 (2012).
\bibitem{jianxing2014} J.-X. Li, K. Z. Hatsagortsyan, and C. H. Keitel, Phys. Rev. Lett. {\bf 113}, 044801 (2014).
\bibitem{krushelnick} Z.-H. He, J. A. Nees, B. Hou, K. Krushelnick, and A. G. R. Thomas,  Phys. Rev. Lett. {\bf 113}, 263904 (2014).

\bibitem{kitagawa} Y. Kitagawa, Y. Mori, O. Komeda, K. Ishii, R. Hanayama, K. Fujita, S. Okihara, T. Sekine, N. Satoh, T. Kurita, M. Takagi, T. Watari, T. Kawashima, H. Kan, Y. Nishimura, A. Sunahara, Y. Sentoku, N. Nakamura, T. Kondo, M. Fujine, H. Azuma, T. Motohiro, T. Hioki, M. Kakeno, E. Miura, Y. Arikawa, T. Nagai, Y. Abe, S. Ozaki, A. Noda,
Phys. Rev. Lett. {\bf 114}, 195002 (2015).
\bibitem{jianxing2015} J.-X. Li, K. Z. Hatsagortsyan, B. J. Galow, and C. H. Keitel,
Phys. Rev. Lett. {\bf 115}, 204801 (2015).

\bibitem{lax} M. Lax, W. H. Louisell, W. B. McKnight,  Phys. Rev. A {\bf 11}, 1365 (1975).

\bibitem{davis} L. W. Davis,  Phys. Rev. A {\bf 19}, 1177 (1979).

\bibitem{barton} J. P. Barton and D. R. Alexander,  J. Appl. Phys. {\bf 66}, 2800 (1989).

\bibitem{esarey} E. Esarey, P. Sprangle, M. Pilloff, and J. Krall, J. Opt. Soc. Am. B {\bf 12}, 1695 (1995).
\bibitem{esarey2} E. Esarey, and W. Leemans,  Phys. Rev. E {\bf 59}, 1082 (1999).
\bibitem{porras} M. A. Porras, Phys. Rev. E {\bf 58}, 1086 (1998).
\bibitem{brabec} T. Brabec and F. Krausz,  Rev. Mod. Phys. {\bf 72}, 545–591 (2000).

\bibitem{winful} S. Feng and H. G. Winful,  Phys. Rev. E, {\bf 61}, 862 (2000).

\bibitem{wang2}  P. X. Wang, and J. X. Wang,  Appl. Phys. Lett. {\bf 81}, 4473 (2002).

\bibitem{lu} D. Lu, W. Hu, Y. Zheng, Z. Yang,  Opt. Commun. {\bf 228}, 217 (2003).

\bibitem{hua} J. F. Hua, Y. K. Ho, Y. Z. Lin, Z. Chen, Y. J. Xie, S. Y. Zhang, Z. Yan, and J. J. Xu,  Appl. Phys. Lett. {\bf 85}, 3705 (2004).

\bibitem{yan} Z. Yan, Y. K. Ho, P. X. Wang, J. F. Hua, Z. Chen, and L. Wu,  Appl. Phys. B {\bf 81}, 813 (2005).

\bibitem{becker} Q. Lin, J. Zheng, and W. Becker, Phys. Rev. Lett. {\bf 97}, 253902 (2006).

\bibitem{sepke_pulse} S. M. Sepke and D. P. Umstadter,  Opt. Lett. {\bf 31}, 2589 (2006).

\bibitem{varin}  C. Varin, M. Pich\'{e}, and M. A. Porras,
J. Opt. Soc. Am. A {\bf 23}, 2027 (2006).

\bibitem{pukhov} D. an der Br\"ugge and A. Pukhov, Phys. Rev. E {\bf 79}, 01663 (2009).

\bibitem{april1} A. April, Opt. Lett. {\bf 33}, 1392 (2008).
\bibitem{april2} A. April, in {\it Coherence and Ultrashort Pulse Laser Emission}; F.J. Duarte, Ed. (InTech: New York, NY, USA), 355 (2010).
\bibitem{gonoskov} I. Gonoskov, A. Aiello, S. Heugel, and G. Leuchs,  Phys. Rev. A {\bf 86}, 053836 (2012).

\bibitem{marceau} V. Marceau, A. April, and M. Pich\'{e},  Opt. Lett. {\bf 37}, 2442 (2012).

\bibitem{wong}  L. J. Wong, F. X. K\"{a}rtner, and S. G. Johnson, Opt. Lett. {\bf 39}, 1258 (2014).

\bibitem{varin1} C. Varin, S. Payeur, V. Marceau, S. Fourmaux,
    A. April, B. Schmidt, P.-L. Fortin, N Thir\'{e}, T. Brabec,
    F. L\'{e}gar\'{e}, J.-C. Kieffer and M. Pich\'{e}, Appl. Sci. {\bf 3}, 70 (2013).
\bibitem{salamin92a} Y. I. Salamin, Phys. Rev. A {\bf 92}, 053836 (2015).
\bibitem{salamin92b} Y. I. Salamin, Phys. Rev. A {\bf 92}, 063818 (2015).
\bibitem{josab33} J.-X. Li, Y. I. Salamin, K. Z. Hatsagortsyan, and C. H. 	Keitel, J. Opt. Soc. Am. B {\bf 33}, 405 (2016).
\bibitem{salamin2010} Y. I. Salamin, Phys. Rev. A {\bf 82}, 013823 (2010).
\bibitem{jianxing2012} J.-X. Li, Y. I. Salamin, B. J. Galow, and C. H. 	Keitel, Phys. Rev. A {\bf 85}, 063832 (2012).
\bibitem{wong2} L. J. Wong and F. X. K\"{a}rtner, Opt. Express {\bf 18}, 25035 (2010).

\bibitem{sell} A. Sell and F. X. K\"{a}rtner, J. Phys. B {\bf 47}, 015601 (2014).

\bibitem{carbajo} S. Carbajo et al., Opt. Lett. {\bf 39}, 2487 (2014).
\bibitem{kaertner} S. Carbajo, E. A. Nanni, L. J. Wong, G. Moriena, P. D. Keathley, G. Laurent, R.-J. D. Miller, and F. X. K\"{a}rtner,
Phys. Rev. Accel. Beams {\bf 19}, 021303 (2016).

\bibitem{mcdonald} \url{http://puhep1.princeton.edu/~kirkmcd/examples/axicon.pdf}

\bibitem{salaminNJP8} Y. I. Salamin, New J. of Phys. {\bf 8}, 133 (2006);
     Y. I. Salamin, New J. of Phys. {\bf 10}, 069801 (2008).


\end{thebibliography}
\end{document}